\documentstyle[prl,aps,epsfig,calc]{revtex}
\begin{document}
\draft

% define commands for international characters
\catcode`\ä = \active \catcode`\ö = \active \catcode`\ü = \active
\catcode`\Ä = \active \catcode`\Ö = \active \catcode`\Ü = \active
\catcode`\ß = \active \catcode`\é = \active \catcode`\è = \active
\catcode`\ë = \active \catcode`\ô = \active \catcode`\ê = \active
\catcode`\ø = \active \catcode`\ò = \active \catcode`\í = \active
\defä{\"a}

\defü{\"u} \defÄ{\"A} \defÖ{\"O} \defÜ{\"U}

\defè{\`{e}}

\defê{\^{e}}

\defí{\'{i}} % preprint mode
% 2 col mode:
\twocolumn[\hsize\textwidth\columnwidth\hsize\csname
@twocolumnfalse\endcsname
\title{Collisions in zero temperature Fermi gases} \vspace{-5mm}
\author{Subhadeep Gupta, Zoran Hadzibabic, James R. Anglin, and Wolfgang Ketterle}
\address{Department of Physics, MIT-Harvard Center for Ultracold
Atoms, and Research Laboratory
of Electronics, \\
MIT, Cambridge, MA 02139}
\date{\today}
\maketitle

\begin{abstract}
We examine the collisional behavior of two-component Fermi gases
released at zero temperature from a harmonic trap. Using a
phase-space formalism to calculate the collision rate during
expansion, we find that Pauli blocking plays only a minor role for
momentum changing collisions. As a result, for a large scattering
cross-section, Pauli blocking will not prevent the gas from
entering the collisionally hydrodynamic regime. In contrast to the
bosonic case, hydrodynamic expansion at very low temperatures is
therefore not evidence for fermionic superfluidity.
\end{abstract}
\pacs{PACS numbers: 03.75.Ss, 03.75.Kk, 34.50.-s} \vskip1pc]

\narrowtext

The last few years have seen rapid progress in the field of
ultracold atomic Fermi gases
\cite{dema99,trus01,schr01,gran02,hadz02,roat02}. Most recently,
regimes of strong interactions have been observed in these gases
near Feshbach resonances \cite{ohar02_2,rega03,bour03,gupt03}.
Studies of these systems are of particular importance because of
the possibility of creating BCS-like superfluids
\cite{houb99,holl01}. Such a realization would establish highly
controllable model systems for studying novel regimes of fermionic
superfluidity.

A unique feature of atomic systems is the ability to analyze the
gas by turning off the trapping potential and observing the
expansion. The expansion behavior can reveal the momentum
distribution and the effects of mean-field interactions and
collisions. Hydrodynamic behavior can be easily detected when the
gas is released from an anisotropic atom trap. In that case, the
spatial anisotropy of the cloud reverses during free expansion.
This is caused by the larger pressure gradient along the tightly
confining direction, which leads to a faster expansion, and
subsequent reversal of the spatial anisotropy. This anisotropic
expansion was used to identify the formation of the Bose-Einstein
condensate \cite{ande95,davi95bec}.

A BEC obeys the hydrodynamic equations of a superfluid
\cite{stri96}. However collisional hydrodynamics arising from a
high elastic collision rate also results in anisotropic expansion
\cite{kaga97,pedr03}. Thus, the normal component can also expand
anisotropically \cite{shva02}. For the bosonic case, two key
points make the distinction between the two fractions obvious: (i)
At the typical transition temperature, the BEC has much less
energy than the normal cloud, so the two components are clearly
separated in size. (ii) The scattering rate needed to achieve
condensation is usually not large enough that the normal gas is in
the hydrodynamic regime. For these two reasons, the appearance of
a dense anisotropic cloud during expansion is considered to be the
``smoking-gun" for the formation of a Bose-Einstein condensate.

A superfluid Fermi gas is predicted to obey the superfluid
hydrodynamic equations of motion
\cite{bara98,ming01,zamb01,meno02} and therefore should show
strong anisotropic expansion when released from an anisotropic
harmonic trap \cite{meno02}. The recent observation of anisotropic
expansion of an ultracold, interacting, two-spin fermionic mixture
\cite{ohar02_2,rega03,bour03} has created considerable excitement
and raised the question under what conditions is this expansion a
signature of fermionic superfluidity and not of collisional
hydrodynamics. There are two major differences from the bosonic
case: (i) Since the energy of ultracold fermions always remains on
the order of the Fermi energy, the size in expansion for both
normal and superfluid components will be similar. (ii) Current
efforts towards inducing BCS pairing all take place in strongly
interacting systems. This results in a large scattering rate
modified only by the effects of Pauli blocking at low
temperatures.

The interpretation of the observed anisotropic expansion in
strongly interacting Fermi gases is therefore critically dependent
on the role of Pauli blocking of collisions during the expansion.
The tentative interpretation of anisotropic expansion as
superfluid hydrodynamics \cite{ohar02_2} was based on the
assessment that collisions are strongly suppressed at sufficiently
low temperatures
\cite{ohar02_2,ferr99,geis99,holl00,vich00,geis02,gehm03}.

Here we show generally that the collision rate becomes independent
of temperature and prevails even at zero temperature, if the Fermi
surface is strongly deformed. This happens in an extreme way
during ballistic expansion. In the small cross-section limit, we
find that less than half of the total number of momentum changing
collisions is suppressed. For a large scattering cross-section,
the absence of suppression results in strong collisional behavior
of normal Fermi gases during expansion for all initial
temperatures. This result has the important consequence of
rendering expansion measurements of Fermi gases near Feshbach
resonances ambiguous for differentiating between superfluid and
normal components.

We first consider the expansion of a single component Fermi gas.
At ultralow temperatures, fermionic antisymmetry prevents s-wave
scattering in a single component and renders the gas completely
collisionless. The phase space occupation
$f(x_1,x_2,x_3,p_1,p_2,p_3)=f({\bf x},{\bf p})$ at zero
temperature in a harmonic trap with frequencies
$(\omega_1,\omega_2,\omega_3)$ can be written as
$$
f({\bf x},{\bf p}) = \Theta(E_F - \Sigma_i m\omega_i^2 x_i^2/2 -
\Sigma_i p_i^2/2m)
$$
where $m$ is the particle mass, $\Theta$ is the Heaviside step
function defined as $\Theta(x)=0(1)$ for $x\leq0(x>0)$ and $E_F=
\hbar (6N \Pi_i \omega_i)^{1/3}$ is the Fermi energy for $N$
particles. At time $t=0$, the trapping potential is turned off
suddenly, allowing the gas to expand freely. At $t=0$, the
momentum space Fermi surface at ${\bf x}=(x_1,x_2,x_3)$ is
\begin{equation}\label{FSintrap}
\Sigma_i p_i^2/2m = E_F - \Sigma_i m\omega_i^2 x_i^2/2,
\end{equation}
a sphere of radius $\sqrt{2m(E_F - \Sigma_i m\omega_i^2
x_i^2/2)}$. In this non-interacting system, the evolution of the
Fermi surface can be derived from the simple evolution law for
ballistic expansion ${\bf x}(0) = {\bf x}(t)- {\bf p}t/m$.
Substituting this in Eq.$\,$\ref{FSintrap}, we obtain:
\begin{equation}\label{FSTOF}
\Sigma_i {\frac{(1 + \omega_i^2 t^2)}{2m}} \left(p_i-{\frac{mx_i
}{t}}{\frac{\omega_i^2 t^2 }{1+ \omega_i^2t^2}}\right)^2 = E_F -
\Sigma_i {\frac{m \omega_i^2 x_i^2 }{{2(1 + \omega_i^2t^2)}}},
\end{equation}
which is an ellipsoid with generally unequal axes $\sqrt{2m(E_F -
\Sigma_i {\frac{m \omega_i^2 x_i^2 }{{2(1 +
\omega_i^2t^2)}}})}({\frac{1}{\sqrt{1+\omega_1^2t^2}}},{\frac{1}
{\sqrt{1+\omega_2^2t^2}}},{\frac{1}{\sqrt{1+\omega_3^2t^2}}})$.
The anisotropy of the Fermi surface during expansion can be
understood generally by noting that for long times $t$, at any
position ${\bf x}$, the spread in momentum $\Delta p_i(t)$ can
only arise from the initial spread in position $\Delta x_i(0)$.
For anisotropic traps this gives rise to an anisotropic momentum
distribution during ballistic expansion. For a mixture of two spin
states, this deformation of the Fermi surface from a sphere into
an anisotropic ellipsoid removes Pauli blocking of final states
and allows collisions, as will be shown.

The momentum distribution at position ${\bf x}$ given by
Eq.$\,$\ref{FSTOF}, also allows us to simply calculate the spatial
density distribution as the volume of the momentum-space
ellipsoid,
\begin{equation}\label{density}
n({\bf x},t)={\frac{4 }{3}}\pi\left({\frac{2mE_F
}{h^2}}\right)^{3/2} {\frac{(1-{\frac{m}{2E_F}} \Sigma_i
{\frac{\omega_i^2 x_i^2 }{1+\omega_i^2 t^2}})^{3/2}} {\Pi_i
(1+\omega_i^2 t^2)^{1/2}}},
\end{equation}
in agreement with other derivations \cite{bruu00}. For long
expansion times $t$, the spatial distribution becomes isotropic,
mirroring the isotropic momentum distribution in the trap.

\begin{figure}[htbf]
\begin{center}
\vskip0mm \epsfxsize=80mm {\epsfbox{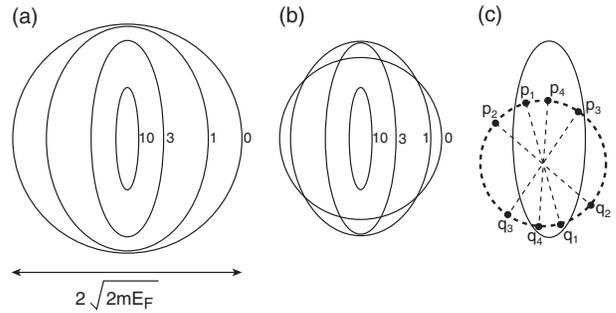}} \vskip1mm
\end{center}
\caption{(a) Deformation of the momentum space Fermi surface at
${\bf x}={\bf 0}$, from a sphere to an ellipsoid during expansion
from an anisotropic harmonic trap. The case of cylindrical
symmetry is shown, where the three-dimensional distribution is
symmetric about the vertical axis. The parameters chosen are an
aspect ratio $\lambda=0.2$ and expansion times $\omega_\perp t =$
0,1,3 and 10. (b) The deformation at a position radially displaced
by $\sqrt{E_F/m\omega^2_\perp}$. (c) Geometrical representation of
collisions in momentum space. The two spin states have identical
distributions. Three different types of collisions are shown for
particles with initial momenta ${\bf p_1}$ and ${\bf q_1}$ - none,
one or both of the final states are occupied. }\label{ellipse}
\end{figure}

Specializing to the experimentally relevant case of a
cylindrically symmetric trap, ballistic expansion deforms the
local Fermi surface into a momentum ellipsoid of cylindrical
symmetry with aspect ratio $\sqrt{{\frac{1+\omega_z^2
t^2}{1+\omega_\perp^2 t^2}}}$ (Fig.$\,$\ref{ellipse}(a,b)). Here
$\omega_\perp (\omega_z)$ is the radial (axial) trapping
frequency. For long times $t$, this deformation reaches the
asymptotic aspect ratio $\omega_z/\omega_\perp=\lambda$, the
initial spatial aspect ratio in the trap.

Now consider an equal mixture of two spin states which interact
via a finite s-wave scattering length. We assume that the trapping
frequencies are identical for the two states (standard
experimental conditions) and specialize to the usual case of
two-body elastic collisions in the local-density approximation.
These collisions have an appealing geometrical picture in the
local phase-space description (Fig.$\,$\ref {ellipse}(c)). Each
elastic collision involves one particle from each spin state. We
label with ${\bf p}$'s and ${\bf q}$'s the momenta of the two
different spin states. Consider the collision ${\bf p_1} + {\bf
q_1} \rightarrow {\bf p_2} + {\bf q_2}$. Conservation of momentum
and kinetic energy mandates ${\bf p_2} + {\bf q_2} = {\bf p_1} +
{\bf q_1}$ and $|{\bf p_2} - {\bf q_2}| = |{\bf p_1} - {\bf
q_1}|$. These relations restrict ${\bf p_2}$ and ${\bf q_2}$ to
lie on diametrically opposite ends of the sphere with ${\bf p_1} -
{\bf q_1}$ as a diameter. The deformation of the Fermi surface
during expansion opens up unoccupied final states ${\bf p_2,q_2}$
and therefore allows collisions to take place even in a zero
temperature Fermi gas (Fig.$\,$\ref {ellipse}(c)).

\vskip5mm

\vskip-3mm

The effect of collisions can be formally calculated from the
Boltzmann transport equation for the evolution of the phase space
distribution $f({\bf x},{\bf p},t)$. In the absence of external
potentials and neglecting mean field we have \cite{huan87}:
\begin{equation}\label{Boltzmann}
{\partial f \over \partial t} + {\bf v}.{\partial f \over
\partial {\bf x}} = \Gamma_{\rm coll}[f]
\end{equation}
where ${\bf v}={\bf p}/m$ and $\Gamma_{\rm coll}[f]$ describes the
effect of collisions. Collisions attempt to restore local
equilibrium by countering the deformation of the momentum space
Fermi surface during free expansion (Eq.$\,$\ref{FSTOF}).

$\Gamma_{\rm coll}[f]$ can be written as the collision integral:
\begin{eqnarray}\label{collint}
&\Gamma({\bf x},{\bf p_1},t)=-{\frac{\sigma }{4\pi h^3}}
\int_{({\bf x},t)}d^3q_1 d^2\Omega {\frac{|{\bf p_1}-{\bf
q_1}|}{m}}\times
\nonumber\\
&[f({\bf p_1})f({\bf q_1})(1-f({\bf p_2}))(1-f({\bf q_2}))
\nonumber\\
&-f({\bf p_2})f({\bf q_2})(1-f({\bf p_1}))(1-f({\bf q_1}))]
\end{eqnarray}
where $\sigma$ is the momentum-independent scattering
cross-section, $f({\bf p_i})=f({\bf x},{\bf p_i},t)$, $f({\bf
q_i})=f({\bf x},{\bf q_i},t)$ and $\Omega$ points along ${\bf p_2}
- {\bf q_2}$. The integral over ${\bf q_1}$ is over the momentum
ellipsoid at position ${\bf x}$ and time $t$ for one of the spin
states. The first term in the integrand is the collision rate for
the process ${\bf p_1} + {\bf q_1} \rightarrow {\bf p_2} + {\bf
q_2}$. The second term corresponds to the reverse process ${\bf
p_2} + {\bf q_2} \rightarrow {\bf p_1} + {\bf q_1}$, and ensures
that only distribution changing collisions contribute.

Pauli blocking is expressed in the suppression factors for the
final states $(1-f)$ in $\Gamma$. The collision integral
neglecting Pauli blocking, $\Gamma_{\rm Cl,p}$, is furnished by
setting the suppression factors all equal to $1$ in
Eq.$\,$\ref{collint}. This is the rate for classical collisions
which change the momentum distribution. The total classical
collision rate $\Gamma_{\rm Cl}$ is the first term on the right
hand side of Eq.$\,$\ref{collint} without any suppression factors.
In addition to $\Gamma_{\rm Cl,p}$, this also contains the rate
for collisions which do not change the momentum distribution: if
both final states are occupied, then the reverse process has the
same rate. These additional collisions do not affect observables
of the system. Fig.$\,$\ref{ellipse}(c) shows examples of these
different types of collisions. ${\bf p_1} + {\bf q_1} \rightarrow
{\bf p_2} + {\bf q_2}$ contributes to $\Gamma_{\rm Cl}$,
$\Gamma_{\rm Cl,p}$ and $\Gamma$. ${\bf p_1} + {\bf q_1}
\rightarrow {\bf p_3} + {\bf q_3}$ contributes to $\Gamma_{\rm
Cl}$ and $\Gamma_{\rm Cl,p}$. ${\bf p_1} + {\bf q_1} \rightarrow
{\bf p_4} + {\bf q_4}$ contributes only to $\Gamma_{\rm Cl}$. To
determine the effect of Pauli blocking, we compare $\Gamma$ and
$\Gamma_{\rm Cl,p}$ for a small cross-section $\sigma$. The
collision rate at a particular time $t$ can then be calculated
perturbatively, by propagating the system ballistically for the
time $t$ and then evaluating Eq.$\,$\ref{collint} with and without
the suppression factors.

\begin{figure}[htbf]
\begin{center}
\vskip0mm \epsfxsize=80mm {\epsfbox{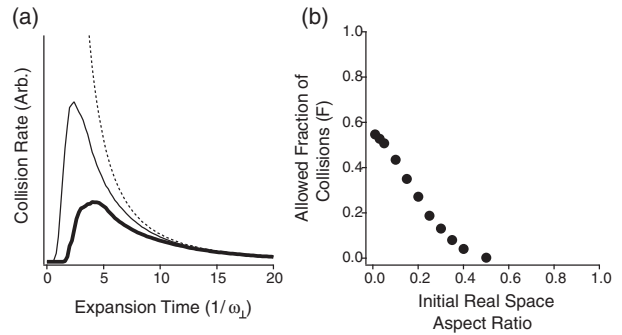}} \vskip1mm
\end{center}
\caption{(a) Collision rate as a function of expansion time in the
perturbative approximation for the initial aspect ratio
$\lambda=0.03$. Dashed line - total classical collision rate
$\Gamma_{\rm cl}$, thin line - classical rate for momentum
changing collisions $\Gamma_{\rm cl,p}$, thick line - collision
rate for fermions $\Gamma$. The displayed rates were evaluated at
${\bf x=0}$ and ${\bf p=0}$ and give an effective upper bound on
the Fermi suppression. (b) Allowed fraction of collisions
$F(\lambda)$ for a zero-temperature two-spin Fermi gas. For an
initial aspect ratio $\lambda=0.05$, $F$ is $0.5$. For large
anisotropy ($\lambda \rightarrow 0$), $F$ approaches $\sim
0.55$.}\label{collrate}
\end{figure}

Fig.$\,$\ref{collrate}(a) displays the numerically calculated
collision rates $\Gamma$, $\Gamma_{\rm Cl,p}$ and $\Gamma_{\rm
Cl}$, evaluated at ${\bf x}\,=\,{\bf p}\,=\,{\bf 0}$, as a
representative case, for an initial aspect ratio $\lambda=0.03$.
Both $\Gamma$ and $\Gamma_{\rm Cl,p}$ increase initially as the
deformation of the Fermi surface becomes more pronounced. For long
times ($\omega_\perp t \gg 1$), they are both suppressed because
both the density ($\int d^3q_1$) and the relative velocity
(${\frac{|{\bf p_1}-{\bf q_1}|}{m}}$) drop. The two curves
approach each other with time since Pauli blocking becomes less
effective with stronger deformation. The fraction of momentum
changing collisions which are not affected by Pauli blocking,
$F(\lambda)=\int dt \Gamma({\bf 0},{\bf 0},t)/\int dt \Gamma_{\rm
Cl,p}({\bf 0},{\bf 0},t)$, is shown in Fig.$\,$\ref{collrate}(b).
The main result of our paper is the inefficiency of Pauli blocking
during expansion from anisotropic traps. For $\lambda<0.05$,
$F>0.5$, and approaches $\sim 0.55$ as $\lambda$ approaches $0$.
Most experiments work in this regime of trap aspect ratio.

The above results form an upper bound on the Fermi suppression
even if we consider all the possible collisions occurring in the
system, for arbitrary ${\bf x}$ and ${\bf p}$. First, we observe
that for all ${\bf x}$, at any time $t$, the Fermi surface is
identically deformed and different only in size according to the
local density (Eqs.$\,$\ref{FSTOF},\ref{density},
Fig.$\,$\ref{ellipse}(a,b)). We have checked numerically  that to
within $5\%$, the central momentum provides a lower bound on
$\Gamma$ within a momentum ellipsoid, at all ${\bf x}$ and for all
times $t$. Next, we note that for ${\bf x}\neq{\bf 0}$, the
density $n({\bf x},t)$ puts more weight at longer times than
$n({\bf 0},t)$ (Eq.$\,$\ref{density}). Since Fermi suppression
becomes less effective with time, Pauli blocking is most effective
at ${\bf x}={\bf 0}$. The calculation for ${\bf x}={\bf p}={\bf
0}$ thus provides an effective upper bound for the overall
collisional suppression in the system. We conclude that more than
half of all the possible collisions are not Pauli blocked for
typical experimental values of $\lambda$.

\vskip-3mm

So far, we have not considered the effect of the collisions
themselves on the momentum distribution. Collisions drive the
system towards equilibrium, which corresponds to an isotropic
Fermi-Dirac distribution. If this collisional relaxation
(Eq.$\,$\ref{collint}) is much faster than the non-equilibrium
perturbation due to ballistic expansion (Eq.$\,$\ref{FSTOF}), the
momentum distribution maintains local equilibrium at all times. If
local equilibrium is maintained, the Boltzmann equation leads to
the hydrodynamic equations \cite{huan87}. For free expansion from
anisotropic atom traps, these equations lead to the reversal of
anisotropy \cite{kaga97,pedr03}. Even if equilibrium is not fully
maintained, collisions always have the effect of transferring
momentum from the weakly confining axis to the strongly confining
axis resulting in an eventual spatial aspect ratio $>1$
\cite{pedr03}.

We now want to reconcile our new result that Pauli blocking is
inefficient during free expansion, with previous results
\cite{vich00} which show that at low temperatures, collisional
damping of collective excitations is suppressed. For this, we
derive an equation of motion for the momentum space anisotropy
$\alpha$ to leading order in $\alpha$ and $T/T_F$ \cite{angl03}:
\begin{equation}\label{alphaeq}
\dot{\alpha}={1\over 3}(\partial_x v_x +\partial_y v_y-2\partial_z
v_z) -{n\sigma p_F\over m}\,C\left(\alpha,{T\over T_F}\right)
\end{equation}
where $C$ describes the collisional relaxation and has the
asymptotic forms:
\begin{eqnarray}\label{alphaeq2}
C\left(\alpha,{T\over T_F}\right)&=&{3\pi^2\over5} \left({T\over
T_F}\right)^2\alpha, \hskip3mm \alpha \ll \left({T\over
T_F}\right)
\nonumber\\
&&{96\over 49}\alpha^3, \hskip16.5mm \alpha \gg \left({T\over
T_F}\right).
\end{eqnarray}
In terms of $\alpha$, the aspect ratio of the momentum space
ellipsoid is $\sqrt{1-\alpha \over 1+2\alpha}$. $p_F,T,T_F$ are
the local Fermi momentum, temperature and Fermi temperature
respectively. This equation was derived from the second momentum
moment of the Boltzmann equation
(Eqs.$\,$\ref{Boltzmann},\ref{collint}), using a Fermi-Dirac
distribution with an anisotropic Fermi surface as ansatz
\cite{corr}. The numerical coefficients in Eq.$\,$\ref{alphaeq2}
were obtained by analytic integrations over momentum space.

At zero temperature, there is no linear term in $\alpha$ in
Eq.$\,$\ref{alphaeq2}. This shows that Pauli blocking is efficient
as long as the anisotropy is small. This is the case for small
amplitude excitations in a {\it trapped} degenerate gas
\cite{vich00}. However, for the large anisotropies of ballistic
expansion, the $\alpha^3$ term, which is independent of
temperature and not affected by Pauli blocking, is responsible for
collisional relaxation.

Eqs.$\,$\ref{alphaeq},\ref{alphaeq2} allow us to distinguish
collisionless from hydrodynamic behavior in different regimes. The
driving term involving ${\bf v}$ is on the order of the trap
frequency $\omega_\perp$ and the damping term has a prefactor
$n\sigma v_F$. Therefore, the dimensionless parameter
characterizing the attainment of the hydrodynamic limit is
$\Phi_0=n\sigma v_F/\omega_\perp$. If $\Phi_0\ll 1$, then one can
neglect collisions entirely, and the gas will expand
ballistically. For small anisotropies, hydrodynamic behavior
requires $\Phi_0 (T/T_F)^2\gg 1$. For large anisotropies,
hydrodynamic behavior requires $\Phi_0^{1/3} \gg 1$. At ultralow
temperatures, the expansion after release from a highly
anisotropic trap may be collisionless initially, but as $\alpha$
grows, the $\alpha^3$ term in Eq.$\,$\ref{alphaeq2} will become
important, and induce hydrodynamic behavior.

Our calculations clearly predict that for parameters of current
experiments, $\Phi_0>1$, free expansion will not be collisionless,
but show behavior which is at least intermediate between
collisionless and hydrodynamic \cite{pedr03}. Full hydrodynamic
behavior may not be achieved, since for small values of $\alpha$,
Pauli suppression becomes effective again. More quantitative
studies are necessary in order to assess how much this behavior
would differ from superfluid expansion. This could be realized by
extending analytical studies \cite{pedr03} to high degeneracies or
by Monte-Carlo techniques \cite{wuar98}. Our main conclusion is
clear, however, that the breakdown of Pauli blocking under free
expansion means that hydrodynamic expansion will not be the
dramatic, qualitative signal for superfluidity in strongly
interacting fermions, the way it was for BEC.

We thank John Thomas, Sandro Stringari, Martin Zwierlein and Aaron
Leanhardt for valuable discussions, and Claudiu Stan and Christian
Schunck for critical reading of the manuscript. This work was
supported by the NSF, ONR, ARO, and NASA.


\vspace{-0.7cm}
\begin{thebibliography}{10}

\bibitem{dema99}
B. DeMarco and D.S. Jin, Science {\bf 285}, 1703 (1999).

\bibitem{trus01}
A.G. Truscott et. al., Science {\bf 291}, 2570 (2001).

\bibitem{schr01}
F. Schreck et. al., Phys. Rev. Lett. {\bf 87}, 080403 (2001).

\bibitem{gran02}
S.R. Granade, M.E. Gehm, K.M. O'Hara, and J.E. Thomas, Phys. Rev.
Lett. {\bf 88}, 120405 (2002).

\bibitem{hadz02}
Z. Hadzibabic et. al., Phys. Rev. Lett. {\bf 88}, 160401 (2002).

\bibitem{roat02}
G. Roati, F. Riboli, G. Modungo, and M. Inguscio, Phys. Rev. Lett.
{\bf 89}, 150403 (2002).

\bibitem{ohar02_2}
K.M. O'Hara et. al., Science {\bf 298}, 2179 (2002).

\bibitem{rega03}
C.A. Regal and D.S. Jin, Phys. Rev. Lett. {\bf 90}, 230404 (2003).

\bibitem{bour03}
T. Bourdel et. al., arXiv:cond-mat/0303079 (2003).

\bibitem{gupt03}
S. Gupta et. al., Science {\bf 300}, 1723 (2003).

\bibitem{houb99}
M. Houbiers and H.T.C. Stoof, Phys. Rev. A. {\bf 59}, 1556 (1999).

\bibitem{holl01}
M. Holland, S.J.J.M.F. Kokkelmans, M.L. Chiofalo, and R. Walser,
Phys. Rev. Lett. {\bf 87}, 120406 (2001).

\bibitem{ande95}
M.H. Anderson et. al., Science {\bf 269}, 198 (1995).

\bibitem{davi95bec}
K.B. Davis et. al., Phys. Rev. Lett. {\bf 75}, 3969 (1995).

\bibitem{stri96}
S. Stringari, Phys. Rev. Lett. {\bf 77},  2360  (1996).

\bibitem{kaga97}
Y. Kagan, E.L. Surkov, and G.V. Shlyapnikov, Phys. Rev. A {\bf
55}, R18 (1997).

\bibitem{pedr03}
P. Pedri, D. Gu$\acute{e}$ry-Odelin, and S. Stringari,
arXiv:cond-mat/0305624 (2003).

\bibitem{shva02}
I. Shvarchuck et. al., Phys. Rev. Lett. {\bf 89}, 270404 (2002).

\bibitem{bara98}
M.A. Baranov and D.S. Petrov, Phys. Rev. A {\bf 58}, R801 (1998).

\bibitem{ming01}
A. Minguzzi and M.P. Tosi, Phys. Rev. A {\bf 63}, 023609 (2001).

\bibitem{zamb01}
F. Zambelli and S. Stringari, Phys. Rev. A {\bf 63}, 033602
(2001).

\bibitem{meno02}
C. Menotti, P. Pedri, and S. Stringari, Phys. Rev. Lett. {\bf 89},
250402 (2002).

\bibitem{ferr99}
G. Ferrari, Phys. Rev. A. {\bf 59}, R4125 (1999).

\bibitem{geis99}
W. Geist et. al., Phys. Rev. A. {\bf 61}, 013406 (1999).

\bibitem{holl00}
M. Holland, B. DeMarco, and D.S. Jin, Phys. Rev. A. {\bf 61},
053610 (2000).

\bibitem{vich00}
L. Vichi, J. Low Temp. Phys. {\bf 121},  177  (2000).

\bibitem{geis02}
W. Geist and T.A.B. Kennedy, Phys. Rev. A {\bf 65}, 063617 (2002).

\bibitem{gehm03}
M.E. Gehm, S.L. Hemmer, K.M. O'Hara, and J.E. Thomas,
arXiv:cond-mat/0304633(2003). The authors point out that the
suppression of in-trap collisions is not sufficient to guarantee
collisionless expansion.

\bibitem{bruu00}
G.M. Bruun and C.W. Clark, Phys. Rev. A {\bf 61},061601 (2000).

\bibitem{huan87}
K. Huang, {\it Statistical Mechanics}, (J. Wiley, New York 1987),
2nd. ed.

\bibitem{angl03}
J.R. Anglin et. al., in preparation (2003).

\bibitem{corr}
The use of the Fermi-Dirac form neglects possible effects of
strong interatomic correlations. Although such effects are not
well understood, they are likely to shift a normal degenerate
Fermi gas even further away from the collisionless limit.

\bibitem{wuar98}
H. Wu and E. Arimondo, Europhys. Lett. {\bf 43},  141  (1998).

\end{thebibliography}
\end{document}